\documentstyle[12pt,fleqn]{article}
\font\st=eusm10 scaled 1200
\begin{document}
\noindent{\large\bf A hypothesis concerning ``quantal''
Hilbert space criterion of chaos in nonlinear dynamical
systems}\\

\vspace{1.5cm}
\noindent Krzysztof Kowalski\\

\vspace{1.5cm}
\noindent {\it Department of Theoretical Physics, University of \L\'od\'z,
ul.\ Pomorska 149/153, 90-236 \L\'od\'z, Poland}
\vspace{1.5cm}

\begin{quote}
\hspace*{1.5em}Based on the Hilbert space approach to the theory
of nonlinear \hbox{dynamical} systems developed by the author a hypothesis
is formulated concerning the ``quantal'' criterion for classical
ordinary differential systems to exhibit chaotic behaviour.
\end{quote}
\vspace{1.5cm}

\vspace{1.5cm}
\newpage
\noindent {\bf 1. Introduction}
\vspace{.8cm}

As so many natural laws and models of natural phenomena are 
described by nonlinear dynamical systems $d{\bf x}/dt={\bf 
F}({\bf x})$, it is no exaggeration to say that they abound
in modern science and technology.  Physical and numerical 
experiments show that deterministic chaos in such systems is 
ubiquitous.  On the other hand, the recent observations 
concerning possibility of controlling chaos [1] indicate its 
new promising applications in engineering and medical 
sciences [2].  No wonder that the interest in the theory of 
chaotic dynamical systems is steadily increasing.

The object of the present paper is to formulate a hypothesis 
concerning ``quantal'' Hilbert space criterion of chaos in 
nonlinear systems $d{\bf x}/dt={\bf F}({\bf x})$, where {\bf 
F} is analytic in {\bf x}, with the use of the formalism 
developed by the author [3-9], relying on reduction of 
nonlinear dynamical systems to the linear, abstract, 
Schr\"odinger-like equation in Hilbert space.  Namely, it is 
suggested that the chaotic behaviour of the nonlinear 
dynamical systems can be related to the growth of the 
``quantal'' entropy which can be naturally introduced within 
the Hilbert space approach resembling quantum mechanics.

\vspace{1.2cm}
\noindent {\bf 2. Hilbert space approach}

\vspace{.8cm}
In this section we briefly outline the Hilbert space 
description of nonlinear ordinary differential systems [3].  
Consider the analytic system (complex or real)
\begin{equation}
\frac{d{\bf z}}{dt} = {\bf F}({\bf z}),\qquad {\bf 
z}(0)={\bf z}_0,
\end{equation}
\noindent where {\bf F}: ${\bf C}^k\to{\bf C}^k$ is analytic 
in {\bf z}.

The vectors $|z,t\rangle$ defined by
\begin{equation}
|z,t\rangle = \exp[\hbox{$\frac{1}{2}$}(|{\bf z}|^2-|{\bf 
z}_0|^2)]|{\bf z}\rangle,
\end{equation}
\noindent where $|{\bf z}\rangle$ is a normalized coherent
state (see appendix) and {\bf z} fulfils (1),  satisfy the 
following linear, Schr\"odinger-like equation in Hilbert 
space:
\begin{equation}
\frac{d}{dt}|z,t\rangle = M|z,t\rangle,\qquad 
|z,0\rangle=|{\bf z}_0\rangle,
\end{equation}
\noindent where $M$ is the boson operator such that
\begin{equation}
M = {\bf a}^\dagger \mbox{\boldmath${\cdot}$}{\bf F}({\bf 
a}).
\end{equation}
\noindent In view of (2) we arrive at the following 
eigenvalue equation:
\begin{equation}
{\bf a}|{\bf z}_0,t\rangle = {\bf z}({\bf z}_0,t)|{\bf 
z}_0,t\rangle
\end{equation}
\noindent relating the solution ${\bf z}({\bf z}_0,t)$ of 
(1) and the solution $|{\bf z}_0,t\rangle$ of (3).  Thus, it
turns out that the nonlinear dynamical system (1) can be 
cast into the linear, abstract, Schr\"odinger-like equation 
(3).  We note that the algorithm can be immediately extended 
to the case of nonautonomous systems such that the 
corresponding vector field is analytic in {\bf z}-variables.  
We also remark that the eigenvalue equation (5) suggests the 
``quantization scheme'' of the form ${\bf z}\to{\bf a}$, 
where {\bf z} fulfils (1), within introduced formalism 
resembling quantum mechanics.

Observe that (3) and (5) correspond to the ``Schr\"odinger 
picture'' within the actual ``quantal'' treatment.  We end 
this brief account of the Hilbert space approach with 
discussion of the ``Heisenberg picture''.  The ``Heisenberg 
equations of motion'' obeyed by the time-dependent Bose 
annihilation operators are of the form
\begin{equation}
\frac{d{\bf a}}{dt} = [{\bf a},M],\qquad {\bf a}(0)={\bf a},
\end{equation}
\noindent where $M$ is the ``Hamiltonian'' given by (4).
Since the ``Hamiltonian'' in ``Schr\"odinger picture'' 
$M$ is independent of time, therefore it coincides with 
the ``Hamiltonian'' in ``Heisenberg picture''.

The formal solution of (6) can be written as
\begin{equation}
{\bf a}(t) = V(t)^{-1}{\bf a}V(t),
\end{equation}
\noindent where $V(t)=e^{tM}$ is the evolution operator.
\noindent It is clear that the following relation holds true:
\begin{equation}
|{\bf z}_0,t\rangle = V(t)|{\bf z}_0\rangle,
\end{equation}
\noindent where $|{\bf z}_0,t\rangle$ is the solution of
(3).

Now eqs.\ (5), (8) and (7) taken together yield
\begin{equation}
{\bf a}(t)|{\bf z}_0\rangle = {\bf z}({\bf z}_0,t)|{\bf
z}_0\rangle.
\end{equation}
\noindent Hence we find that the solutions of the systems
(1) coincide with the expectation values (covariant symbols)
of the time-dependent Bose annihilation operators, i.e.
\begin{equation}
{\bf z}({\bf z}_0,t) = \langle {\bf z}_0|{\bf a}(t)|{\bf
z}_0\rangle.
\end{equation}
\noindent It should be noted that whenever the
``quantization scheme'' ${\bf z}\to{\bf a}$ is assumed,
where {\bf z} fulfils (1), then (10) forms the ``Ehrenfest's
theorem'' within the actual ``quantal'' approach.

\vspace{1.2cm}
\noindent {\bf 3. The criterion of chaos}

\vspace{.8cm}
We now formulate the hypothesis concerning the criterion for
the nonlinear dynamical systems (1) to show chaotic
behaviour.  Consider the system (1).  We introduce the
operator $\rho$ of the form (see (A.2)):
\begin{equation}
\rho = \int\limits_{{\bf R}^{2k}} d\mu({\bf w})\,e^{-|{\bf w}-{\bf
z}_0|^2}|{\bf w}\rangle\langle {\bf w}|.
\end{equation}
\noindent Evidently, the operator $\rho$ is Hermitian and
nonnegative, i.e.
$$\displaylines{
\hspace{2.5em} \rho^\dagger = \rho,\hfill\llap{(12a)}\cr
\hspace{2.5em} \rho\ge0.\hfill\llap{(12b)}\cr}$$
\noindent Furthermore, using (A.6) we immediately obtain
$$\displaylines{\hspace{2.5em}
\hbox{Tr}\rho=1.
\hfill\llap{(12c)}\cr}$$
\noindent Finally, taking into account (11), (A.2), (A.5) 
and (10) we get
$$\displaylines{\hspace{2.5em}
\langle {\bf a}(t)\rangle = \hbox{Tr}(\rho{\bf a}(t)) = {\bf z}(t),
\hfill\llap{(12d)}\cr}$$
\noindent where ${\bf a}(t)$ are the time-dependent Bose 
annihilation operators and ${\bf z}(t)$ fulfils (1).

We have thus shown that if the ``quantization scheme'' ${\bf 
z}\to{\bf a}$, where {\bf z} satisfies (1) is assumed, then 
$\rho$ plays the role of the density matrix within the
``quantal'' Hilbert space approach.  Prompted by this 
analogy, we define the ``entropy'' as
\begin{equation}
\setcounter{equation}{13}%
S = -\hbox{Tr}(\rho\ln\rho).
\end{equation}
\noindent On using the relation
\begin{equation}
\hbox{Tr}\rho^n = \frac{1}{(2^n-1)^k},\qquad n\ge1,
\end{equation}
\noindent following directly from (11), (A.2) and (A.4), we 
find
\begin{equation}
S = -\sum_{i=0}^{\infty}{i+k-1\choose
i}\frac{1}{2^{i+k}}\ln\frac{1}{2^{i+k}} =
2k\ln2.
\end{equation}
\noindent Since an arbitrary vector in a Fock space can be
specified by an infinite sequence of $n$-vectors of 1's
and 0's and the number of different $n$-vectors is $2^n$, 
therefore the obtained value of the ``entropy'' can be
regarded as an averaged amount of the information necessary 
to fix the coherent state $|{\bf z}_0\rangle$ in a Hilbert 
space of states.

Now in analogy to quantum mechanics, we introduce the 
time-dependent ``density matrix'' such that
\begin{equation}
\rho(t) = \int\limits_{{\bf R}^{2k}} d\mu({\bf w})\,e^{-|{\bf w}-{\bf
z}_0|^2}e^{tM}|{\bf w}\rangle\langle {\bf w}|e^{tM^\dagger },
\end{equation}
\noindent where $M$ is the``Hamiltonian'' related to the 
system (1).

We note that $\rho(t)$ is Hermitian and
nonnegative at any time.  The time-dependent ``entropy'' 
corresponding to (16) is given by
\begin{equation}
S(t) = -\hbox{Tr}[\rho(t)\ln\rho(t)].
\end{equation}
\noindent We are now in a position to propound our hypothesis.
We assert that if the following conditions hold:
$$\displaylines{
\hspace{2em} \mathop{\hbox{$\exists$}}\limits_{t_*>0}
\,\mathop{\hbox{$\forall$}}\limits_{t>t_*}|{\bf
z}({\bf z}_0,t)|^2<C,\hfill\llap{(18a)}\cr}$$
\noindent where ${\bf z}({\bf z}_0,t)$ is the solution to
(1); $C>0$ is a constant,
$$\displaylines{
\hspace{2em}\mathop{\hbox{$\forall$}}\limits_{t>t_*}\frac{dS(t)}{dt}>0,
\hfill\llap{(18b)}\cr}$$
\noindent then the system (1) is chaotic one.

\vspace{1.2cm}
\noindent {\bf 4. Discussion}

\vspace{.8cm}
We have formulated in this work a hypothesis concerning a 
new, ``quantal'' Hilbert space criterion of chaos in 
nonlinear dynamical systems.  Note that the condition (18a) 
ensures that the ``density matrix'' $\rho(t)$ is of
trace class for all $t>t_*$.  Indeed, from (16) and the
relation
\begin{equation}
\setcounter{equation}{19}%
\langle {\bf z}_0,t|{\bf z}_0,t\rangle = \exp(|{\bf z}({\bf
z}_0,t)|^2-|{\bf z}_0|^2),
\end{equation}
\noindent which is an immediate consequence of (2), it
follows that
\begin{equation}
\hbox{Tr}\rho(t) = \int\limits_{{\bf R}^{2k}}
d\mu({\bf w})\exp(-|{\bf w}-{\bf
z}_0|^2)\exp(|{\bf z}({\bf w},t)|^2-|{\bf w}|^2),
\end{equation}
\noindent where ${\bf z}({\bf z}_0,t)$ is the solution to
(1).

On the other hand, the condition (18a) is consistent with 
the exceptional role of dissipativity of systems showing 
chaotic behaviour.  We remark that the nonunitary evolution 
implied by the Schr\"odinger-like equation (3) is crucial 
for the actual treatment.  In fact, as in quantum mechanics, 
the unitary evolution implies $S(t)=S={\rm const}$.  We note 
at the same time that the evolution
operator $V(t)=e^{tM}$ defined by (8) is unitary only in the
case of the linear system (1) with a skew-Hermitian
fundamental matrix.  Finally, we point out that the ``quantal''
Hilbert space approach recognizes the divergence of orbits in
the phase space of the system (1).  Namely, we have the
formula
\begin{equation}
\frac{|\langle {\bf z}_0,t|{\bf z}_0+\delta{\bf
z}_0,t\rangle|^2}{\langle {\bf z}_0,t|{\bf
z}_0,t\rangle\langle {\bf z}_0+\delta{\bf z}_0,t|{\bf
z}_0+\delta{\bf z}_0,t\rangle} = \exp(-|{\bf z}({\bf
z}_0+\delta{\bf z}_0,t)-{\bf z}({\bf z}_0,t)|^2)
\end{equation}
\noindent which can be easily obtained from (2) and (A.1).
Therefore, the growth of the distance between the two
trajectories of the system (1) starting from ${\bf z}_0$ and
${\bf z}_0+\delta{\bf z}_0$ implies the
decay of the corresponding transition probability from the
left-hand side of (21).

\vspace{1.2cm}
\noindent {\bf Appendix. Coherent states}

\vspace{.8cm}
We append the basic facts about coherent states.  Recall
first that the Bose creation (${\bf a}^\dagger $) and
annihilation ({\bf a}) operators, where ${\bf a}^\dagger
=(a_1^\dagger ,\ldots ,a_k^\dagger )$, ${\bf a}=(a_1,\ldots
,a_k)$, obey the Heisenberg algebra
$$\displaylines{
\hspace{2.5em}[a_i,a_j^\dagger] = \delta_{ij}I, \hfill\cr
\hspace{2.5em}[a_i,a_j]=[a_i^\dagger ,a_j^\dagger ]=0,\qquad i,\,j=1,\,\ldots
\,,\,k.\hfill\cr}$$
\noindent The coherent states $|{\bf z}\rangle$, where ${\bf
z}\in{\bf C}^k$, are usually defined as eigenvectors of the
annihilation operators, that is
\begin{displaymath}
{\bf a}|{\bf z}\rangle = {\bf z}|{\bf z}\rangle.
\end{displaymath}
\noindent The normalized coherent states can
be defined as
\begin{displaymath}
|{\bf z}\rangle = \exp(\hbox{$-\frac{1}{2}$}|{\bf z}|^2)
\exp({\bf z}\mbox{\boldmath${\cdot}$}{\bf a}^\dagger )|{\bf 0}\rangle.
\end{displaymath}
\noindent where $|{\bf z}|^2=\sum_{i=1}^{k}|z_i|^2$, ${\bf
u}\mbox{\boldmath${\cdot}$}{\bf v}=\sum_{i=1}^{k}u_iv_i$ and
$|{\bf 0}\rangle$ is the vacuum vector satisfying
\begin{displaymath}
{\bf a}|{\bf 0}\rangle = {\bf 0}.
\end{displaymath}
\noindent The coherent states are not orthogonal, namely
\begin{displaymath}
\langle {\bf z}|{\bf w}\rangle =
\exp[-\hbox{$\frac{1}{2}$}(|{\bf z}|^2+|{\bf w}|^2-2{\bf
z}^*\mbox{\boldmath${\cdot}$}{\bf w})],
\end{displaymath}
\noindent where the asterisk designates the complex
conjugation and ${\bf z}^*=(z_1^*,\ldots ,z_k^*)$.

Hence we find
$$\displaylines{
\hspace{2.5em} |\langle {\bf z}|{\bf w}\rangle|^2 =
\exp(-|{\bf z}-{\bf w}|^2).\hfill\llap{(A.1)}\cr}$$
\noindent These states form the complete (overcomplete) set.  The
resolution of the identity for the coherent states is given 
by
$$\displaylines{
\hspace{2.5em} \int\limits_{{\bf R}^{2k}}d\mu({\bf z})\,|{\bf
z}\rangle\langle {\bf z}| = I,
\hfill\llap{(A.2)}\cr}$$
\noindent where
\begin{displaymath}
d\mu({\bf z}) =
\prod_{i=1}^{k}\frac{1}{\pi}d(\hbox{Re}\,z_i)\,d(\hbox{Im}\,
z_i).
\end{displaymath}
\noindent Now let $|\phi\rangle$ be an arbitrary state.  It
can be easily shown that the function (symbol of the vector)
$\phi({\bf z}^*)=\langle {\bf z}|\phi\rangle$ is of the form
$$\displaylines{
\hspace{2.5em} \phi({\bf z}^*) = \tilde \phi({\bf
z}^*)\exp(-\hbox{$\frac{1}{2}$}|{\bf z}|^2),
 \hfill\llap{(A.3)}\cr}$$
\noindent where $\tilde \phi({\bf z}^*)$ is an analytic
(entire) function.

Taking into account (A.2) and (A.3) we get
\begin{displaymath}
\langle \phi|\psi\rangle = \int\limits_{{\bf
R}^{2k}}d\mu({\bf z})\exp(-|{\bf z}|^2)(\tilde \phi({\bf
z}^*))^*\tilde\psi({\bf z}^*).
\end{displaymath}
\noindent Thus the abstract vectors can be represented by
analytic functions.  This representation is usually known as
the Bargmann representation.  The action of the Bose
operators in the Bargmann representation has the following
form:
\begin{displaymath}
{\bf a}\tilde \phi({\bf z}^*) = \frac{\partial}{\partial
{\bf z}^*}\tilde \phi({\bf z}^*),\qquad
{\bf a}^\dagger \tilde \phi({\bf z}^*) = {\bf z}^*\tilde
\phi({\bf z}^*).
\end{displaymath}
\noindent On using (A.2) and (A.3) we arrive at the
following reproducing property of coherent states:
$$\displaylines{
\hspace{2.5em} \tilde \phi({\bf w}^*) = 
\int\limits_{{\bf R}^{2k}}d\mu({\bf
z})\exp(-|{\bf z}|^2)\hbox{\st\symbol{75}}({\bf w}^*,{\bf z})
\tilde \phi({\bf z}^*),
 \hfill\llap{(A.4)}\cr}$$
\noindent where the reproducing kernel (Bergman reproducing 
kernel) is
\begin{displaymath}
\hbox{\st\symbol{75}}({\bf w}^*,{\bf z}) = \exp({\bf
w}^*\mbox{\boldmath${\cdot}$}{\bf z}).
\end{displaymath}
\noindent Taking the Hermitian conjugate of (A.4) we
obtain the following form of the reproducing property:
$$\displaylines{\hspace{2.5em}
\tilde \psi({\bf w}) = \int\limits_{{\bf R}^{2k}}d\mu({\bf
z})\exp(-|{\bf z}|^2)\exp({\bf
w}\mbox{\boldmath${\cdot}$}{\bf z}^*)\tilde \psi({\bf z}).
\hfill\llap{(A.5)}\cr}$$
\noindent We finally remark that the coherent states are the
convenient tool for the study of operators.  For example,
the trace of a linear operator $L$ can be expressed by
$$\displaylines{
\hspace{2.5em} \hbox{Tr}L = \int\limits_{{\bf 
R}^{2k}}d\mu({\bf z})\,L({\bf 
z}^*,{\bf z}),
 \hfill\llap{(A.6)}\cr}$$
\noindent where
\begin{displaymath}
L({\bf z}^*,{\bf z}) = \langle {\bf z}|L|{\bf z}\rangle
\end{displaymath}
\noindent is called the covariant symbol of the operator $L$
[10].
\newpage
\noindent {\bf References}

\vspace{.8cm}
\noindent [1]\quad \parbox[t]{14cm}{E. Ott, C. Grebogi and
J.A. Yorke, Phys. Rev. Lett. 64 (1990) 1196.}\\[.2cm]
\noindent [2]\quad \parbox[t]{14cm}{A. Garfinkel, M.L.
Spano, W.L. Ditto and J.N. Weiss, Science 257 (1992) 1230.}\\[.2cm]
\noindent [3]\quad \parbox[t]{14cm}{K. Kowalski, Physica A
145 (1987) 408.}\\[.2cm]
\noindent [4]\quad \parbox[t]{14cm}{K. Kowalski, Physica A
152 (1988) 98.}\\[.2cm]
\noindent [5]\quad \parbox[t]{14cm}{K. Kowalski and W.-H.
Steeb, Progr. Theor. Phys. 85 (1991) 713.}\\[.2cm]
\noindent [6]\quad \parbox[t]{14cm}{K. Kowalski and W.-H.
Steeb, Progr. Theor. Phys. 85 (1991) 975.}\\[.2cm]
\noindent [7]\quad \parbox[t]{14cm}{K. Kowalski and W.-H.
Steeb, Nonlinear Dynamical Systems and Carleman
Linearization (World Scientific, Singapore, 1991).}\\[.4cm]
\noindent [8]\quad \parbox[t]{14cm}{K. Kowalski, Physica A
195 (1993) 137.}\\[.2cm]
\noindent [9]\quad \parbox[t]{14cm}{K. Kowalski, Physica A
198 (1993) 493.}\\[.2cm]
\noindent [10]\quad \parbox[t]{14cm}{F.A. Berezin, Mat. Sb.
86 (1971) 578 (Russian).}

\end{document}